\documentstyle [12pt,epsf]{article}

\input{epsf}
\begin{document}
\begin{titlepage}
\begin{center}

\vspace{-0.7in}

{\large \bf Effective Lagrangians for Scalar Fields\\
and Finite Size Effects in Field Theory}\\
\vspace{.4in}{\large\ M.I.Caicedo\footnotemark[1]\footnotemark[2] and
N.F.Svaiter\footnotemark[3]\footnotemark[4]}\\

Center for Theoretical Physics,\\
Laboratory for Nuclear Physics and Department of Physics,\\
Massachusetts Institute of Technology\\
Cambridge, Massachusetts 02139 USA

\subsection*{\\Abstract}
\end{center}

We discuss the approach of effective field theory on
a $d$-dimensional Euclidean space in a scalar theory with two 
different mass scales in the presence of flat surfaces.
Then considering Dirichlet and Neumann boundary conditions, 
we implement the renormalization program in the $\lambda\varphi^{4}$ theory 
in a region bounded by two parallel hyperplanes in the one-loop approximation.

\footnotetext[1]{e-mail:mcaicedo@lns.mit.edu}
\footnotetext[2]{On leave from Universidad Simon Bolivar}
\footnotetext[3]{e-mail:svaiter@lns.mit.edu}
\footnotetext[4]{On leave from Centro Brasileiro de Pesquisas Fisicas-CBPF}

PACS numbers:03.70+k,04.62.+v

\end{titlepage}
\newpage\baselineskip .37in

\section{Introduction}

The quantum field theory of self-interacting scalar fields has long served as
a laboratory for developing methods of analysis that can be applied to 
theories of more direct physical interest. 
Our purpose is to investigate the Euclidean field 
theory of two interacting scalar fields
where after the construction of an effective theory for the 
light modes, in some limit, there is a decoupling between the light and heavy modes
as stated by the Appelquist-Carrazzone theorem \cite{appel}. 
Aditionally we impose boundary conditions on the resulting theory in 
order to study finite-size effects and 
the renormalization program in systems where the translational 
symmetry is broken.  
The interest of the literature in the 
study of quantum fields in the presence of boundaries appears after the   
problem investigated by Casimir more than fifty years ago \cite{casimir}.
A complete review of this effect can be 
found in refs. \cite{Greiner} \cite{col} \cite{most}.

In 1948, Casimir showed that neutral perfectly conducting
parallel plates in vacuum attract each other, 
the effect can be interpreted as follows: the distortion of the vacuum 
fluctuations of the quantized electromagnetic field due to the presence 
of metallic plates renders the zero-point energy of 
the field into a measurable quantity.
In the absence of any classical 
background, the renormalized vacuum expectation 
value of the Hamiltonian operator
can be correctly defined by 
the Wick-ordered product. 
The main support for this procedure comes from the proof that for a 
relativistic field theory, the vacuum expectation value of the stress-energy 
tensor should vanish in order to ensure that the realization of the Poincar\'e 
generators in terms of the fields of the theory satisfy the correct 
commutation relations \cite{takahashi}. On the other hand, where  external 
fields or macroscopic structures are present, more elaborated methods must be used 
to find the renormalized vacuum 
energy of the quantized field while avoiding undesirable divergences.
A procedure that identifies the divergent contributions to the vacuum energy, 
such as a cut-off or any 
analytic regularization method, followed by a 
renormalization is mandatory. The fundamental idea of the Casimir 
renormalization procedure is that
although formally divergent, the 
difference between the zero-point energies of 
two different physical configurations 
can be shown to be finite.

In quantum electrodynamics there is a standard 
argument used to support the implementation of a 
regularization procedure followed by renormalization
to obtain the renormalized vacuum 
expectation value of the stress-energy tensor 
associated with the Maxwell field in the presence of boundaries.
At high frequencies no real material is a perfect conductor.
A wavelength cut-off corresponding to the finite plasma 
frequency must be included in the model.
High energy modes are insensitive to the boundaries and only the low energy 
modes are affected by them. 
Consequently, in the study of quantum 
electrodynamics in the presence of conducting boundaries,
in the generating functional for the n-point correlation 
functions this latter
condition means that one may integrate out all the Fourier modes associated 
with the Dirac field and obtain an effective theory for the Maxwell field.

We would like to implement the above discussed situation 
in a very simple model of two 
Euclidean self-interacting scalar 
fields. We will analyse a theory with two massive scalar fields with 
different mass scales and we will be interested in obtaining the effective action 
for the light field. 
One could theoretically envision a theory with two massive fields with 
different mass scales on which the mass of the heavy field is smaller than the 
natural cut-off of the boundary, nevertheless in this paper we are not 
interested to discuss 
this situation.
It is important to keep in mind that
in order to construct an effective action that gives a correct 
description of the physics of the light modes 
in the presence of the boundaries, the Fourier modes  
associated with both fields with wavelength smaller 
than $\Lambda^{-1}$ for
some cut-off $\Lambda$ must be integrated out,
restricting the space of functions that we are integrating over 
(note that we are always assuming a sharp cut-off).
Since dealing with functional integrals with cut-offs in general models 
is quite complicated, we will limit ourselves to a heavy field with Gaussian 
functional integrals.
We are studying finite size effects for the light modes 
in two steps. In the first, we integrate over 
the modes of the heavy field, obtaining an effective action for the light field. 
In the second step we are taking the limit in which
the decoupling theorem is valid ($m_{2}\rightarrow \infty$), to regard 
the effective action as the fundamental action and 
we are assuming boundary conditions over the remaining light field. 
A combination of different analytic regularization 
procedures and a renormalization 
are able to eliminate the the usual 
bulk and also the additional surface divergences
that appear in the theory.  
The final result of our procedure is that
we have the effect of the 
compactification of one dimension, 
breaking the full translational invariance of the 
original theory.
In this situation we have an effective field theory of the light modes, 
with finite size effects. Of course, the region outside the boundaries is the 
union of two simple connected domains and the renormalization of the interacting 
field theory in such region must be carried out along the  
same lines that we have to use in the interior region. For simplicity we are only 
focussing our analysis to the interior region.

There are many papers in the literature 
discussing quantum field theory in the presence of 
boundaries or macroscopic structures. 
The calculation of the radiative corrections to the renormalized 
energy density associated with the Maxwell field, in 
the presence of perfectly conducting plates,  
assuming no boundary conditions for the Dirac field 
was performed by Bordag et al \cite{bb}. Temperature corrections 
for this model were analyzed by 
Robaschik et al \cite{rr}. Bordag et al also studied the 
leading radiative correction to the renormalized energy, assuming that 
the paralel plates are represented by delta function potentials \cite{bbb}.
Using the approach of effective field theory, the radiative correction to the 
Casimir effect integrating out the fermionic degrees of freedom was examined by
Kong and Ravndal and Ravndal and Thomassen  \cite{kong}. 
A different approach was used by Falomir et al. \cite{falomir}. These
authors  studying scalar fields in the presence of a spherical shell
used a sharp cut-off, 
assuming that the boundary is transparent for the heavy modes while
for the soft modes they assumed Dirichlet boundary conditions. 
 Actor \cite{actor} studied two interacting scalar 
fields in the presence of 
macroscopic boundaries assuming that 
only one of the fields satisfies classical 
boundary conditions. Using the generalized zeta function 
method \cite{gen} the 
one-loop effective action was presented. 
More recently, Melnikov \cite{melnikov} 
investigated the low-energy effective action 
in a model with two scalar fields and also in quantum electrodynamics.
Our treatment is very similar to the treatment developed by Melnikov in the 
study of the low-energy effective action for a theory with two interacting 
scalar fields.

It is important to point out that the combination of effective field
theory and finite size effects can produce unexpected new
phenomena. A well-known example of this situation is the
Scharnhorst effect \cite{sch}. 
Studying quantum
electrodynamics between perfectly conducting plates, Scharnhorst concluded
that the speed of light normal to the plates, exceeds light speed
on vacuum, while parallel to the plates light travels
with its vacuum speed, i.e., with that of light propagating in
unbounded space. Further calculations by Barton \cite{bar1} and
also by Barton and Scharnhorst \cite{bar2} confirmed the original
result. The Scharnhorst effect is probably a
consequence of the combined use of the effective Lagrangian in
quantum electrodynamics and the presence of the plates, i.e., the
Euler-Heisenberg Lagrangian density \cite{eu}, derived many years
ago and also rederived later by Schwinger \cite{sc}. 
As we already pointed out, the central issue of this effective Lagrangian 
is a derivative
expansion of the photon effective action obtained by integrating
out the fermionic field in the Maxwell-Dirac action.
For an interesting discussion 
concerning the velocity of propagation of signals in 
different field theory models, see for example, \cite{papa}.

Finite size effects that do not break translational 
invariance in quantum field theory also have been 
extensively studied in the literature 
\cite{bi} \cite{npb}.
For translationally invariant systems, we can change from coordinate
space to momentum space representation, the latter being a more
convenient framework to analyze the divergences of the $n$-point
Schwinger functions on which translational invariance is realized
through conservation conditions. For systems where the
translational invariance has been partially broken (so there is
still translational invariance along certain directions) a more
convenient representation for the $n$-point Schwinger functions 
is a mixed momentum-coordinate representation. Important 
references discussing the renormalization 
program in the presence of boundaries are the Symanzik \cite{sy}
and also Diehl and Dietrich \cite{diel} papers.

In this paper we are studying the renormalization program in the 
presence of surfaces where a scalar  
field satisfies boundary conditions. 
We are interested in investigating a very simple
model where we can construct an effective field theory for the light field 
on which the decoupling theorem holds. 
Further we consider an interacting
Euclidean field theory of the light field in the presence of boundaries.
We will consider a Casimir-like
configuration where one of the coordinates, $z$, lies in the
interval $[0,L]$ imposing Dirichlet-Dirichlet boundary conditions
and for the sake of
completeness we will also enquire about Neumann-Neumann boundary
conditions.

The organization of the paper is as follows: In section
II we introduce a simple model of two Euclidean
interacting scalar fields. By 
integrating out the heavy modes associated with one of the 
fields, we are able to 
build the effective action for the remaining field. In section
III we discuss a scalar field
theory with boundary conditions due to the
finite size of one of the coordinates and build the free two-point
and four-point functions, both for Dirichlet-Dirichlet and
Neumann-Neumann boundary conditions. In section IV we
discuss the surface divergences of the one-loop two-point 
and also four-point function
function. In section V we discuss the global approach, used to define the Casimir 
energy associated with a field in the presence of boundaries with 
well defined geometric shape.. 
Finally, section VI contains our conclusions. Throughout this paper we use $\hbar=c=1$.

\section{The Euclidean functional integral and the 
effective action}

The goal of this section is to present 
a very simple model of two Euclidean self-interacting 
scalar fields where after the construction of 
an effective action \cite{wu} \cite{mano} \cite{db} \cite{dobado},
and further imposition of the infinite mass limit for the heavy field,
the decoupling theorem holds. 
For this purpose we  start from a model 
with two different mass scales.  
We consider two real interacting massive scalar fields
$\varphi_{1}(x)$ and $\varphi_{2}(x)$ with masses $m_{1}$ and
$m_{2}$, respectively and regard the 
field $\varphi_2(x)$
as the heavy field ($m_{2}>>m_{1}$). 
For this theory,
we will rederive the Euclidean version for the Appelquist-Carrazonne theorem 
stating that for certain choice for the self-interacting $\varphi_{1}(x)$ part
and in the limit
$m_{2}\rightarrow\infty$,
there is a decoupling in the effective theory.
The only effects of the heavy field
$\varphi_{2}(x)$ being a  modification of the value of the renormalized
mass and the coupling constant of the light field $\varphi_{1}(x)$.

We start from the generating   
functional for the n-point Schwinger functions associated with 
two massive real fields 
in a $d$-dimensional Euclidean space given by
\begin{equation}
Z[j_{1},j_{2}]={\cal N} 
\int [d\varphi_{1}] [d\varphi_{2}]\,e^{-S[\varphi_{1},\varphi_{2}]+(source\,terms)},
\label{nn}
\end{equation}
where $[d\varphi_{1}][d\varphi_{2}]=
\prod_{x\in R^{d}} d\varphi_{1}(x)d\varphi_{2}(x)$ is an appropriate measure,  
$S[\varphi_{1},\varphi_{2}]$ is the classical action associated with 
the Euclidean fields, and in the 
generating functional, ${\cal N}$ is a normalization. 
As usual, the n-point Schwinger functions of the theory  
can be obtained by functional differentiation with respect to the external  
sources $j_{1}(x)$ and $j_{2}(x)$.  Since in this section our 
interest is to construct the effective theory for the light field,  
the introduction of the external sources in the functional integral 
is not important for our discussion. 
We consider the theory described by the following Euclidean
Lagrangian density  with two real scalar fields
\begin{equation}
{\cal L}(\varphi_{1},\varphi_{2})={\cal
L}_{0}(\varphi_{1},\varphi_{2})
+{\cal L}_{int}(\varphi_{1},\varphi_{2}),
\label{1}
\end{equation}
where the free part of the Euclidean Lagrangian density is given by
\begin{equation}
{\cal
L}_{0}(\varphi_{1},\varphi_{2})=\frac{1}{2}(\partial_{\mu}\varphi_{1})^{2}+
\frac{1}{2}m_{1}^{2}\varphi_{1}^{2}+
\frac{1}{2}(\partial_{\mu}\varphi_{2})^{2}+
\frac{1}{2}m_{2}^{2}\varphi_{2}^{2},
\label{2}
\end{equation}
and the interacting part is given by
\begin{equation}
{\cal
L}_{int}(\varphi_{1},\varphi_{2})=V(\varphi_{1})+\frac{\lambda_{2}}{2}
(\varphi_{1}\varphi_{2})^{2}.
\label{3}
\end{equation}
It is important to remark that 
the precise form of $V(\varphi_{1})$ is not
important for the construction of 
the effective action, but as we will see later in this section,   
the form of the $V(\varphi_{1})$ is important to implement the 
decoupling theorem.

The action
of the model is given by
\begin{equation}
S[\varphi_{1},\varphi_{2}]=\int d^{d}x\, {\cal
L}(\varphi_{1}(x),
\varphi_{2}(x)),
\label{4}
\end{equation}
and using Eq.(\ref{2}) and Eq.(\ref{3}) can be conveniently split up as
\begin{equation}
S[\varphi_{1},\varphi_{2}]=S[\varphi_{1}(x)]
+S_{\varphi_{2}}[\varphi_{1}(x),\varphi_{2}(x)],
\label{5}
\end{equation}
$S[\varphi_{1}(x)]$ being the $\varphi_{2}(x)$-independent part of
it. In order to obtain a derivative expansion of the effective
action $\Gamma_{eff}[\varphi_{1}]$ we have to assume $m_{2}>>m_{1}$.
As usual, the operators $(-\Delta+m^{2}_{1})^{-1}$ and
$(-\Delta+m^{2}_{2})^{-1}$ must be used to define the free
two-point Schwinger functions of the fields $\varphi_{1}(x)$ and $\varphi_{2}(x)$
respectively, and we must recall that $\Delta$ stands for the
Laplacian operator in ${\cal R}^{d}$. According to the above, the free
two-point Schwinger functions of both fields can be represented  by
\begin{equation}
G(x-y;m_{i})=<x|(-\Delta+m^{2}_{i})^{-1}|y>, \,\,{i=1,2}
\label{6}
\end{equation}
and they obviously satisfy
\begin{equation}
(-\Delta+m^{2}_{i})G(x-y;m_{i})=\delta^{d}(x-y). 
\label{8}
\end{equation}
To obtain an effective action for the light
modes of the theory, we integrate out the heavy field $\varphi_{2}(x)$
in the functional integral and define
the effective action of the light modes $\Gamma_{eff}[\varphi_{1}]$ by 
\begin{equation}
e^{-\Gamma_{eff}[\varphi_{1}]}=
\int [d\varphi_{2}]\,e^{-S[\varphi_{1},\varphi_{2}]}.
\label{9}
\end{equation}
Using Eq.(\ref{5}) it is possible to write Eq.(\ref{9}) as
\begin{equation}
e^{-\Gamma_{eff}[\varphi_{1}]}
=e^{-S[\varphi_{1}]}\int[d\varphi_{2}]\, e^{-
S_{\varphi_{2}}[\varphi_{1},\varphi_{2}]}.
\label{11}
\end{equation}
The first step in our calculation is straightforward, since   
we have
\begin{equation}
S_{\varphi_{2}}[\varphi_{1},\varphi_{2}]=\int d^{d}x
\,(\varphi_{2}(-\Delta+m^{2}_{2})\varphi_{2}+
\frac{\lambda_{2}}{2}(\varphi_{1}\varphi_{2})^{2}), 
\label{12}
\end{equation}
and using Eq.(\ref{12})
the functional integral  
appearing in Eq.(\ref{11}) can be performed by
means of Gaussian integrations, yielding
\begin{equation}
e^{-\Gamma_{eff}[\varphi_{1}]}=e^{-S[\varphi_{1}]}(det\,
O)^{-\frac{1}{2}},
\label{13}
\end{equation}
where we have 
\begin{equation}
O(x,y;m_{2})=<x|O|y>=(-\Delta_{x}+
m_{2}^{2}+\lambda_{2}\varphi_{1}^{2}(x))\delta^{d}(x-y).
\label{14}
\end{equation}
Consequently, the effective action for the light 
field $\varphi_{1}(x)$ is given by
\begin{equation}
\Gamma_{eff}[\varphi_{1}]=S[\varphi_{1}]+\frac{1}{2}tr\,ln\,
O.
\label{15}
\end{equation}
Dropping a term that contributes trivially to the effective
action $\Gamma_{eff}[\varphi_{1}]$, we get  
\begin{equation}
\Gamma_{eff}[\varphi_{1}]=S[\varphi_{1}]+
\frac{1}{2}tr\,ln
(1+\lambda_{2}(-\Delta_{x}+m^{2}_{2})^{-1}\varphi_{1}^{2}).
\label{17}
\end{equation}
There are many ways to evaluate the Fredholm determinant, defined by 
the above equation. Using a series expansion,
it is possible to rewrite Eq.(\ref{17}) as
\begin{equation}
\Gamma_{eff}[\varphi_{1}]=S[\varphi_{1}]+
\frac{1}{2}\sum_{k=1}^{\infty}\frac{(-1)^k}{k}tr(
\lambda_{2}(-\Delta_{x}+m^{2}_{2})^{-1}\varphi_{1}^{2}))^{k},
\label{18}
\end{equation}
or in a more compact notation  
\begin{equation}
\Gamma_{eff}[\varphi_{1}]=S[\varphi_{1}]+\sum^{\infty}_{k=1}
\Gamma^{(k)}[\varphi_{1}],
\label{19}
\end{equation}
where each term of the series $\Gamma^{(k)}[\varphi_{1}]$
is given by Eq.(\ref{18}).
Let us study the first non trivial contribution of this series,
namely, the term $k=1$ which in fact corresponds to a one-loop
diagram. It is explicitly given by
\begin{equation}
\Gamma^{(1)}[\varphi_{1}]=
\frac{\lambda_{2}}{2}\int d^{d}x\,
G(x-x;m_{2})\,\varphi_{1}^{2}(x).
\label{20}
\end{equation}
Using the Fourier representation for the 
two-point Schwinger function associated with 
the heavy field $\varphi_{2}(x)$, and defining a
new coupling constant $\sigma=\lambda_{2}\mu^{4-d}$,
where $\mu$  is the usual dimensional parameter that appears in the dimensional 
regularization procedure, we readily obtain
\begin{equation}
\Gamma^{(1)}[\varphi_{1}]=\frac{\sigma}{2(2\sqrt{\pi})^{d}}
\Gamma(1-\frac{d}{2})
(m_{2})^{d-2}
\int d^{d}x~\varphi_{1}^{2}(x).
\label{21}
\end{equation}
Now, the Gamma function $\Gamma(z)$ is a
meromorphic function of the complex variable $z$ with simple poles
at the points $z=0,-1,-2,..$
In the neighborhood of any of
its poles $z=-n$, for $n=0,1,2,..$ $\Gamma(z)$ has a
representation given by
\begin{equation}
\Gamma(z)=\frac{(-1)^{n}}{n!}\frac{1}{(z+n)}+\Omega(z+n),
\label{gam}
\end{equation}
where $\Omega(z+n)$ stands for the regular part of the analytic
extension of $\Gamma(z)$. Note that for odd dimensions  
$\Gamma^{(1)}[\varphi_{1}]$
is completely
regular while in even dimensions ($d=2\ell$  $\ell=1,2,...$) there
are singularities in the dimensional regularized quantity.
By using the standard dimensional
regularization prescription $d=2\ell-\epsilon$ and  
since the Eq.(\ref{21}) is quadratic in the field $\varphi_{1}(x)$,
the divergence can be absorbed in the renormalized
$\varphi_{1}(x)$ mass. Consequently, we define the renormalized
mass for the light field as
\begin{equation}
m^{2}_{1R}=m^{2}_{1}+\frac{\sigma{}m_{2}^{d-2}}{(4\pi)^{\frac{d}{2}}}
\left[\frac{(-1)^{\frac{d}{2}-1}}{(\frac{d}{2}-1)!}\frac{2}{\epsilon}+
\Omega(\frac{\epsilon}{2})\right]. \label{eft1}
\end{equation}
For the odd dimensional case there are no poles,
but we have the same situation.
We have thus shown that  
at the one-loop approximation, the first correction to the effective action 
given by $\Gamma^{(1)}[\varphi]$
that we obtain integrating out the heavy field
$\varphi_{2}(x)$ is just a modification of the value of the
renormalized mass associated with the light field. 

We will now show that the second correction to the effective
action given by $\Gamma^{(2)}[\varphi_1]$ only modifies the value of
the coupling constant of the field $\varphi_{1}(x)$. To this end,
let us study the second term of the series in Eq.(\ref{19}). It
corresponds to a one-loop diagram and is actually given by
\begin{equation}
\Gamma^{(2)}[\varphi_{1}]=
\frac{\sigma^{2}}{4}\int d^{d}x \int d^{d}y\, G(y-x;m_{2})
G(x-y;m_{2})\varphi_{1}^{2}(x) \varphi_{1}^{2}(y),
\label{25}
\end{equation}
which by upon substitution of the free two-point Schwinger function 
associated with the $\varphi_{2}(x)$ heavy field 
and the introduction of  $I(p^{2},m^{2}_{2})$ as
\begin{equation}
I(p^{2},m_{2}^{2})= \frac{1}{(2\pi)^{d}}
\int d^{d}q 
\frac{m^{4-d}_{2}}{(q^{2}+m_{2}^{2})((p+q)^{2}+m_{2}^{2})},
\label{98}
\end{equation}
can be written as
\begin{equation}
\Gamma^{(2)}[\varphi_{1}]=\frac{\sigma^{2}}{4(2\pi)^{d}}
\int d^{d}x \int d^{d}y
\,\varphi_{1}^{2}(x) \varphi_{1}^{2}(y)
\int d^{d}p\, e^{-ip(y-x)}m_{2}^{d-4}
I(p^{2},m_{2}^{2}).
\label{97}
\end{equation}
In the regularization and renormalization procedure we have to
eliminate the poles and their residues adding counterterms in the
Lagrangian density, consequently let us study the $I(p^{2},m_{2}^{2})$. Using
the Feynman parametrization \cite{to}, it is possible to write  $I(p^{2},m_{2}^{2})$ as
\begin{equation}
I(p^{2},m_{2}^{2})=
N_{d}(-\frac{1}{\epsilon}+O(\epsilon))
(1-\frac{d}{2})
\int^{1}_{0}dt(\frac{p^{2}}{m^{2}_{2}}t(1-t)+1)^{\frac{d}{2}-2},
\label{92}
\end{equation}
where $N_{d}$ is the area of the sphere
$S_{d-1}/(2\pi)^{d}$.
The expression given by Eq.(\ref{92}) contains a power of a
binomial in a flat Euclidean 
d-dimensional space. When $d$ is even, the
power is an integer and the simple use of Newton's binomial
theorem will give us a very direct way of evaluating $I(p^{2},m_{2}^{2})$.
When $d$ is odd, the  expansion of
$(1+\frac{p^{2}}{m^{2}_{2}}t(1-t))^{\frac{d}{2}-2}$
yields an infinite power series. Since we are using dimensional 
regularization we have an infinite power series. 
Note that the generalization of the 
binomial series is valid for any complex exponent $p$. In other words
we have an everywhere convergent power series in $p$, hence a continuous function
on p in the complex plane  \cite{ss}. 
If we define
\begin{equation}
C^{0}_{r}=1,\quad{}C^{1}_{r}=
\frac{r}{1!},\quad{}C^{2}_{r}=\frac{r(r-1)}{2!}\dots
\label{91}
\end{equation}
\noindent{}until
\begin{equation}
C_{r}^{k}=\frac{r(r-1)..(r-k+1)}{k!},
\label{90}
\end{equation}
where $r=\frac{d}{2}-2$, it is possible to write
$I(p^{2},m_{2}^{2})$ as
\begin{equation}
I(p^{2},m_{2}^{2})=(1-\frac{d}{2})N_{d}(-\frac{1}{\epsilon}+O(\epsilon))
\sum_{k=0}^{\infty}C_{\frac{d}{2}-2}^{k}\frac{p^{2k}}{m_{2}^{2k}}
\int^{1}_{0}dt(t(1-t))^{k}.
\label{89}
\end{equation}
Let us use the definition of Euler's integral of first kind
$B(\alpha,\beta)$ given by \cite{grads}.
\begin{equation}
B(\alpha,\beta)=\int_{0}^{1} dx\, x^{\alpha-1}(1-x)^{\beta-1}=
\frac{\Gamma(\alpha)\Gamma(\beta) }{\Gamma(\alpha+\beta) }, \,\, Re\,\alpha>0,
Re\,\beta>0.
\label{88}
\end{equation}
Substituting Eq.(\ref{89}) and Eq.(\ref{88}) in Eq.(\ref{97})
we find
that the second term of the series that represents the  
effective action, $ \Gamma^{(2)}[\varphi_{1}]$ can be written as
\begin{eqnarray}
\Gamma^{(2)}[\varphi_{1}]& &=\frac{\sigma}{4(2\pi)^{d}}
\int d^{d}x \int d^{d}y
\,\varphi_{1}^{2}(x) \varphi_{1}^{2}(y)
\int d^{d}q \,\,
e^{-ip(y-x)}m_{2}^{d-4}\nonumber\\
&&((1-\frac{d}{2})N_{d}(-\frac{1}{\epsilon}+O(\epsilon))
\sum_{k=0}^{\infty}C_{\frac{d}{2}-2}^{k}\frac{p^{2k}}{m_{2}^{2k}}
B(k+1,k+1).
\label{86}
\end{eqnarray}
If we choose the self-interacting part of the field
$\varphi_{1}(x)$ to
be $\lambda_{1}\varphi_{1}^{4}(x)$, it is possible to
define the renormalized coupling constant $\lambda_{R}$ subtracting the
polar part.
Consequently, the effective action for the $\varphi_{1}(x)$ field
is given by
\begin{eqnarray}
\Gamma_{eff}[\varphi_{1}]&=&\frac{1}{2}
\int d^{d}x
~\varphi_{1}(x)(-\Delta+m_{1}^{2})\varphi_{1}(x)+
\lambda_{R} \int d^{d}x ~\varphi_{1}^{4}(x)\nonumber\\
& &~~~+\frac{\sigma^{2}}{4!(4\pi)^{2}m_{2}^{2}}
\int d^{d}x~
\varphi_{1}^{2}(x)(-\Delta+m_{1}^{2})\varphi_{1}^{2}(x)+
O(\frac{\Delta}{m^{2}_{2}})^{2}.
\label{28}
\end{eqnarray}
Note that the terms $k=3,4..$ 
are not divergent (in a four dimensional theory) and although they 
contribute to the effective action, 
in the limit $m_{2}\rightarrow\infty$, the
heavy field
$\varphi_{2}(x)$ decouples from the light field
$\varphi_{1}(x)$. The
effect of the heavy field appears only modifying the
values of the
renormalized mass $m_{1R}$ and the coupling constant
$\lambda_{R}$. Thus we showed that the Euclidean 
version of the Appelquist-Carazzone 
decoupling theorem works in this specific model. 
Another well known example 
where the decoupling theorem can be used is in quantum electrodynamics, 
where for energies much lower than the 
electron mass it is possible to construct a derivative expansion of
the Maxwell field effective action integrating out the 
Dirac field. This is an expected result, since we known that 
the theorem is valid for renormalizable theories without 
spontaneous symmmetry breaking or chiral fermions. 
The above discussion justifies the approach used by some  
authors that have been using the
Euler-Heisenberg Lagrangian density to 
investigate the radiative correction to the Casimir 
effect \cite{kong}, although these radiative corrections are of no phenomenological 
significance as was pointed out by Melnikov \cite{melnikov}. 
For a careful discussion of effective Lagrangians in 
quantum electrodynamics, see for example ref. \cite{di}.

It is important to stress that instead of obtain also a effective theory 
for the light field as have been discussed by many authors in 
finite temperature field theory 
\cite{polo},  assuming some ultraviolet cut-off  $\Lambda$
and integrating 
over the Fourier modes with wavelenght smaller than $\Lambda^{-1}$, 
we are assuming that the functional integral must be taken 
over the space of the functions that vanish on the boundaries. 
One way to implement this is to introduce delta functions in the 
functional integral. This is equivalent to evaluate the functional 
integral over a space of functions that satisfy the boundary 
conditions. This is the procedure that we are addopting.
It is clear that 
this procedure will introduce additional 
surface divergences that can be eliminated 
by surface counterterms, and in the end we have the effective model for the 
light modes that satisfies boundary conditions over some surfaces.

\section{Finite size effects and the two and 
four-point Schwinger functions in the one-loop approximation}

In the last section, we studied a very simple model of two Euclidean 
massive scalar fields where the decoupling theorem can be 
used after the construction of an 
effective theory of the light field. 
We have shown that in our model of 
two massive self-interacting scalar 
fields, the heavy modes associated with the $\varphi_{2}(x)$ field 
completely decouple from the light ones 
associated with the light field $\varphi_{1}(x)$  
in the limit $m_{2}\rightarrow\infty$.
In the case of a
$\lambda_{1}\varphi_{1}^{4}(x)$ self-coupling,
the only effect of the
former is a modification of the mass $m_{1}$, 
and the coupling constant of the light
field. We are reducing the problem in this manner   
since we are able to concentrate in such a one
field theory, i.e., we will consider a $\lambda_{1}\varphi^{4}(x)$
self-interacting model. We
will consider that the field $\varphi(x)$ depends on
$d-1$ unbounded coordinates that we call 
$\vec{r}$, and one bounded coordinate to which we
will refer to as $z$ that will be assumed to lie in the interval
$[0,L]$. If we exclude the possibility of periodic or anti-periodic 
boundary conditions,  
this choice obviously breaks the full
translational invariance because we have to assume boundary
conditions on the hyperplanes $z=0$ and $z=L$.

To write the full renormalized action for the theory with boundaries 
we need two regulators, the first one being the usual $\epsilon$ that is introduced in 
the dimensional regularization procedure and the second one that we call $\eta$ representing 
the distance to a boundary. According to this 
the full renormalized action must be given by \cite{diel}:
\begin{eqnarray}
S(\varphi)& &= 
\int_{0}^{L}dz \int d^{d-1}r
(\frac{A(\epsilon)}{2}(\partial_{\mu}\varphi)^{2}+\frac{B(\epsilon)}{2}\varphi^2+
\frac{C(\epsilon)}{4!}\varphi^{4})\nonumber\\
& &+
\int d^{d-1}r(c_{1}(\eta)\varphi^{2}(\vec{r},0)+c_{2}(\eta)
\varphi^{2}(\vec{r},L))\nonumber\\
& &+
\int d^{d-1}r(c_{3}(\eta)\varphi^{4}(\vec{r},0)+c_{4}(\eta)
\varphi^{4}(\vec{r},L),
\label{sur}
\end{eqnarray}
where $A(\epsilon)$, $B(\epsilon)$ and $C(\epsilon)$ are 
the usual coefficients for the bulk counterterms and 
the coefficients $c_{i}(\eta)$ $i=1,..4$, which 
depend on the boundary conditions for the field, 
are the coefficients for the surface 
counterterms. As usual all of these coefficients   
must be calculed order by order 
in  perturbation theory. We are considering two
different possibilities for the boundary conditions, namely
Dirichlet-Dirichlet ($DD$) and Neumann-Neumann ($NN$) boundary conditions. 
These boundary conditions are given respectively by
\begin{equation}
\varphi(\vec{r},z)|_{z=0}=\varphi(\vec{r},z)|_{z=L}=0,
\label{dir}
\end{equation}
and
\begin{equation}
\frac{\partial}{\partial{z}}\varphi(\vec{r},z)|_{z=0}
=\frac{\partial}{\partial{z}}\varphi(\vec{r},z)|_{z=L}=0.
\label{new}
\end{equation}
The system we are
interested is invariant only under translations along the
direction parallel to the plates, implying that what is conserved
is not the full momentum but the $(d-1)$ dimensional parallel
momentum $\vec{p}$. For such conditions, a more convenient
representation for the $n$-point Schwinger functions is a mixed $(\vec{p},z)$
one. A Euclidean scalar field $\varphi(x)$ satisfying
certain homogeneous boundary conditions on $z=0$ and $L$  can be 
expanded in Fourier series as:
\begin{equation}
\varphi(\vec{r},z)=\frac{1}{(2\pi)^{\frac{d-1}{2}}} 
\sum_{n} u_{n}(z)
\int
d^{d-1}p\, \phi_{n}(\vec{p})\, e^{i\vec{p}.\vec{r}},
\label{field}
\end{equation}
where $\vec{p}$ is the 
continuum parallel momentum, and $u_{n}(z)$
stands for the eigenfunctions of the operator
$-\frac{d^2}{dz^{2}}$:
\begin{equation}
-\frac{d}{dz^{2}}u_{n}(z)=k^{2}_{n}u_{n}(z),
\label{ult}
\end{equation}
where $k_{n}=\frac{n\pi}{L}$, $n=1,2..$ for $DD$ b.c 
and 
$n=0,1,2..$ for $NN$ boundary conditions respectively. 
The main difference between both boundary conditions 
being the
presence of the zero mode. The free two-point Schwinger 
function for the theory can be expressed in the following form:
\begin{equation}
G_{0}^{(2)}(\vec{r},z,\vec{{r}^{\prime}},z^\prime)=
\frac{1}{(2\pi)^{d-1}}\sum_{n}
u_{n}(z)u_{n}^{*}(z')
\int d^{d-1}p
\frac{e^{i\vec{p}.(\vec{r}-\vec{{r}^{\prime}})}}{(\vec{p}^{\,}+k_{n}^{2}+m^{2})}.
\label{geex}
\end{equation}
Note that in this section we have changed the notation as 
follows: $m_{1}\rightarrow m$ and 
also $G(x,x';m_{1})\rightarrow G_{0}(x,x')$. 
It is useful to define also $\vec{\rho}=\vec{r}-\vec{{r}^{\prime}}$.
When considering $DD$ boundary conditions,
one finds that the free two-point Schwinger function
is explicitly given by
\begin{equation}
G_{0}^{(2)}(\vec{\rho},z,z^{\prime})=
\frac{2}{L}
\sum_{n=1}^{\infty}
\sin(\frac{n\pi{}z}{L})\sin(\frac{n\pi{}z^{\prime}}{L})
I_{n}(L,m,d,\vec{\rho}),
\label{g1}
\end{equation}
where
\begin{equation}
I_{n}(L,m,d,\vec{\rho})=\frac{1}{(2\pi)^
{d-1}}\int{}d^{d-1}p\frac{e^{i\vec{p}.\vec{\rho}}}
{(\vec{p}^{\,2}+(\frac{n\pi}{L})^{2}+m^{2})}.
\label{v}
\end{equation}
It is clear that the family of $I_{n}(L,m,d,\vec{\rho})$
functions
can be thought of as the free propagators of a tower of massive
scalar fields in $(d-1)$ dimensions, the 
effective mass of each mode
being given by $M_n^2=m^2+(\frac{n\pi}{L})^2$.  This
is to be expected since our theory has been formulated in a
compactified space. From an even simpler point of view,
$I_{n}(L,m,d,\vec{\rho})$
is nothing but the Fourier transform of a
``spherically" ($SO(d-1)$) symmetric function of the parallel
momentum $\vec{p}$.

We begin the study of the interacting theory by building the
one-loop correction ($G_{1}^{(2)}(\lambda_{1},x,x')$) to the bare two-point
Schwinger function $G_{0}^{(2)}(x,x')$, for both the $DD$ and $NN$ 
boundary conditions. Using the Feynman rules we have
\begin{equation}
G_{1}^{(2)}(\lambda_{1},\vec{r}_{1},z_{1},\vec{r}_{2},z_{2})
=\frac{\lambda_{1}}{2}\int{}d^{d-1}r\int_{0}^{L}dz
\,\,G_{0}^{(2)}(\vec{r}_{1}-\vec{r},z_{1},z)
G_{0}^{(2)}(\vec{0},z)
G_{0}^{(2)}(\vec{r}-\vec{r}_{2},z,z_{2}).
\label{MF}
\end{equation}
Here we would like to point out that even though the functions
$G_{0}^{(2)}(\vec{r}_{1}- \vec{r}_{2},z_{1},z_{2})$ and
$G_{0}^{(2)}(\vec{r}_{2}-\vec{r}_{3},z_{2},z_{3})$ are singular at
coincident points ($\vec{r}_{1}=\vec{r}_{2}$, $z_{1}=z_{2}$) and
($\vec{r}_{2}=\vec{r}_{3}$, $z_{2}=z_{3}$), the singularities are
integrable for points outside the plates.
Using the notation $G_{0}^{(2)}(\vec{0},z)=T_{DD}(L,m,d,z)$,
a straightforward substitution yields the order $\lambda_{1}$
correction to the bare two-point Schwinger function in the one-loop
approximation, for the case of Dirichlet boundary conditions:
\begin{eqnarray}
G_{1}^{(2)}(\lambda_{1},\vec{r}_{1}-\vec{r}_{2},z_{1},z_{2})& & =
\frac{2\lambda_{1}}{(2\pi)^{d-1}L^{2}}\int_{0}^{L}dz\sum_{n,n^{\prime}=1}^{\infty}
\sin(\frac{n\pi z_{1}}{L})\sin(\frac{n\pi{}z}{L})\sin(\frac{n^{\prime}\pi{}z}{L})
\sin(\frac{n^{\prime}\pi{}z_{2}}{L})\nonumber \\
& & \int d^{d-1}p
\frac{e^{i\vec{p}(\vec{r}_{1}-\vec{r}_{2})}}
{(\vec{p}^{\,2}+(\frac{n\pi}{L})^{2}+m^{2})
(\vec{p}^{\,2}+(\frac{n^{\prime}\pi}{L})^{2}+m^{2})}
T_{DD}(L,m,d,z),
\label{grande}
\end{eqnarray}
where, since we are using dimensional
regularization techniques, we have introduced a dimensional
parameter $\mu$, defining a dimensionless coupling constant 
$\lambda=\lambda_{1}\mu^{4-d}$, and   
the expression for the amputated 
one-loop two-point function $T_{DD}(L,m,d,z)$ is 
given by
\begin{equation}
T_{DD}(L,m,d,z)=\frac{2}{(2\pi)^{d-1}L}\sum_{n=1}^{\infty}
\sin^{2}(\frac{n\pi{}z}{L})
\int d^{d-1}p \frac{1}
{(\vec{p}^{\,2}+(\frac{n\pi}{L})^{2}+m^{2})}.
\label{TadDD}
\end{equation}
In the case of Neumann-Neumann boundary conditions  
the expression for the amputated one-loop two-point function can also be found
following the same procedure, and it is given by
\begin{eqnarray}
T_{NN}(L,m,d,z)& &=\frac{1}{(2\pi)^{d-1}L}\int 
d^{d-1}k\frac{1}
{(\vec{k}^{2}+m^{2})}\nonumber\\
& &+\frac{2}{(2\pi)^{d-1}L}\sum_{n=1}^{\infty}
\cos^{2}(\frac{n\pi z}{L})
\int d^{d-1}p \frac{1}
{(\vec{p}^{\,2}+(\frac{n\pi}{L})^{2}+m^{2})}.         
\label{TdaNN}
\end{eqnarray}
Both $T_{DD}(L,m,d,z)$ and
$T_{NN}(L,m,d,z)$ diverge in their continuum momenta integrals and
also in the discrete mode summation.
Using the Feynman rules,
$G_{2}^{(4)}(\lambda,x_{1},x_{2},x_{3},x_{4})$, i.e.,  
the $O(\lambda^{2})$
correction to the bare one-loop four-point Schwinger functions, 
is given by
\begin{eqnarray}
G_{2}^{(4)}(\lambda,\vec{r}_{1},z_{1},\vec{r}_{2},z_{2},
\vec{r}_{3},z_{3},\vec{r}_{4},z_{4})
&=& \frac{\lambda^2}{2}\int{}d^{d-1}r\int{}d^{d-1}r^{\prime}
\int_{0}^{L}dz \int_{0}^{L}dz^{\prime}
\;G_{0}^{(2)}(\vec{r}_{1}-\vec{r},z_{1},z)
\nonumber \\
&&G_{0}^{(2)}(\vec{r}_{2}-\vec{r},z_{2},z)
(G_{0}^{(2)}(\vec{r}-\vec{r^{\prime}},z,z^{\prime}))^2\nonumber\\
&& G_{0}^{(2)}(\vec{r^{\prime}}-\vec{r}_{3},z^{\prime},z_{3})
G_{0}^{(2)}(\vec{r^{\prime}}-\vec{r}_{4},z^{\prime},z_{4}).
\label{nova}
\end{eqnarray}
Again, all $G_{0}$'s are singular at coincident points, but the
singularities are integrable for points outside the 
plates, except for
$G_{0}^{(2)}(\vec{r}-\vec{r^{\prime}},z,z^{\prime})$.

In the next section we will begin the
renormalization program for the massless one-loop two-point Schwinger functions 
for the case of Dirichlet-Dirichlet boundary condition. The study of the
complementary set of boundary conditions, namely $NN$ boundary conditions can
be performed along the same lines. 
When the fields satisfy $NN$ boundary 
conditions that infrared 
divergences for massless fields appear, and in fact, such divergences
come from the zero mode contribution, so the two-point Schwinger function
for the case of Dirichlet-Dirichlet boundary conditions
is IR finite for $m=0$. 
For the case of $NN$ we must have a finite Euclidean volume to regularize the 
theory in the infrared. Another way to deal with the infrared divergences in the 
case of Neumann-Neumann boundary condition in to perform a resummation of the 
daisy diagrams  \cite{doja} \cite{ka} \cite{dru} \cite{outros}.
Although this procedure 
is standard in the study of scalar models where the translational invariance 
is maintained, for systems where the translational invariance is broken, 
the problem of how to carry out the resummation program
still remains open.

\section{The regularized one-loop two and 
four-point Schwinger functions}

In this section 
we would like to discuss in detail how to implement the one-loop renormalization 
program in finite size systems where flat surfaces break the 
translational invariance. Thus,
the aim of this section is first to analyze the structure of the divergences
associated with the one-loop two and also four-point function 
for the case of Dirichlet-Dirichlet boundary boundary
conditions.  

The amputated one-loop two-point Schwinger function 
$T_{DD}(L,m,d,z)$ can be decomposed 
in a translational invariance part and another one that 
breaks the translational invariance, indeed using algebraic identities \cite{grads} \cite{ap} one gets
\begin{equation}
T_{DD}(L,m,d,z)=f_1(L,m,d)-f_2(L,m,d,z),
\label{split-TDD}
\end{equation}
where the functions $f_{1}(L,m,d)$ and  $f_{2}(L,m,d,z)$
are given respectivelly by
\begin{equation}
f_1(L,m,d)=
\frac{1}{2(2\pi)^{d-1}L}
\sum_{n=-\infty}^{\infty}
\int d^{d-1}p \frac{1}
{(\vec{p}^{\,2}+(\frac{n\pi}{L})^{2}+m^{2})}
\label{f1}
\end{equation}
and
\begin{equation}
f_{2}(L,m,d,z)=
\frac{1}{2(2\pi)^{d-1}}\int{d}^{d-1}p\frac{1}{\sqrt{\vec{p}^2+m^2}}
\frac{cosh((L-2z)\sqrt{\vec{p}^2+m^2})}{senh(L\sqrt{\vec{p}^2+m^2})}.
\label{f2}
\end{equation}
The amputated one-loop two-point Schwinger function  
for the Neumann-Neumann boundary conditions, 
$T_{NN}(L,m,d,z)$ can be similarly split
up as
\begin{equation}
T_{NN}(L,m,d,z)=f_1(L,m,d)+f_2(L,m,d,z)
\label{split-TNN}
\end{equation}
The above decompositions of $T_{DD}(L,m,d,z)$ and $T_{NN}(L,m,d,z)$ have
the same functional form and  as we stated before,
some of the divergences come purely from the bulk while others
depend on the distance to the boundaries. 
Indeed, since $f_1(L,m,d)$ does not depend on $z$, it only carries 
information about the divergences on the bulk. These divergences occur 
not only in the discrete sums but also in the momentum integrations.
After the identification: $\beta\equiv{}2L$ $f_1(L,m,d)$
is formally proportional to the amputated  
one-loop two-point function of
the theory assuming that the system is in thermal equilibrium with
a reservoir at temperature $\beta^{-1}$.
To deal with the divergences of $f_1(L,m,d)$, or equivalently, the
one-loop two-point Schwinger functions at finite temperature we have to do
frequency sums and $(d-1)$ dimensional integrals for the continuum
momenta. One way to perform the integrals with   
Matsubara sums is to analytic extend
away from the discrete complex energies down to real axis with the 
replacement of the energy sums by contour integrals \cite{kapusta} \cite{kap}. Another 
way is to use dimensional regularization and afterwards to 
analytically extend the modified Epstein zeta function which appears 
after dimensional regularization. Direct use of dimensional regularization 
identities and the analytic extension of the 
modified Epstein zeta function in the
sum given by Eq.(\ref{f1}) which defines $f_{1}(L,m,d)$, 
give us a polar part
(size independent) plus a size-dependent analytic correction. The mass counterterm 
(the principal part of the Laurent series of the analytic regularized 
quantity) generated by $f_{1}(L,m,d)$ is
size independent, because the finite temperature field theory has no
temperature dependent counterterm. 
Observe that the non translational invariant part of the amputated one-loop 
two-point Schwinger function 
expressed by $T_{DD}(L,m,d,z)$ and $T_{NN}(L,m,d,z)$
have the same $z$ dependent part in modulus but with opposite
signs. 

Since we have shown that the 
$T_{DD}(L,m,d,z)$ and $T_{NN}(L,m,d,z)$ can be split into two
functions $f_{1}(L,m,d)$ and $f_{2}(L,m,d,z)$ 
and since as we have just discussed the behavior of $f_{1}(L,m,d)$, we can 
now turn our attention to the 
study of the divergences contained in $f_2(L,m,d,z)$. We begin by
an angular integration
($d^{d-1}p=p^{d-2}dp\,d\Omega_{d-1}$ and,  
$\int{}d\Omega_d=\frac{2\pi^{\frac{d}{2}}}{\Gamma(\frac{d}{2})}$ )
that leads to an alternate expression for the non translational invariant 
part $f_2(L,m,d,z)$, namely
\begin{equation}
f_2(L,m,d,z)=
\frac{1}{2}h(d)\int_{0}^{\infty}dp\frac{{p}^{d-2}}{\sqrt{p^2+m^2}}
\frac{cosh((L-2z)\sqrt{p^2+m^2})}{sinh(L\sqrt{p^2+m^2})}.
\end{equation}
Using the change of
variables $s=\sqrt{p^2+m^2}$ in the above expression  yields the following formula
for $f_{2}(L,m,d,z)$:
\begin{equation}
f_2(L,m,d,z)=\frac{1}{2}h(d)\int_{m}^{\infty}ds(s^2-m^2)^{\frac{d-3}{2}}
cosh((L-2z)s)(\sinh Ls)^{-1},
\label{eq:f2-integrado-en-angulo}
\end{equation}
where $h(d)$ is an analytic function of $d$ given by $h(d)=\frac{1}{2(2{\sqrt \pi})^{d-1}}\frac{1}
{\Gamma(\frac{d-1}{2})}$.
We now start studying the massless case following 
Fosco and Svaiter \cite{fo}. In fact, we are particularly
interested in examining the limits ($z\rightarrow{}0^+$ and
$z\rightarrow{}L^-$) which do obviously contain the information
about the effects of the boundaries. In order to fulfill this goal
we introduce two new variables $x=Ls$ and $q=zs$ in terms of which
we can write $f_2(L,m,d,z)|_{m=0}$ as
\begin{eqnarray}
f_2(L,m,d,z)|_{m=0}&=&\frac{h(d)}{2L^{d-2}}\int_{0}^{\infty}dx\,x^{d-3}
(\coth x -1)\cosh(\frac{2zx}{L})
\nonumber\\
&+&\frac{h(d)}{2z^{d-2}}\int_{0}^{\infty}dq\,q^{d-3}e^{-2q}.
\label{ult2}
\end{eqnarray}
The second term of Eq.(\ref{ult2}) give us 
the well known result that for a massless
scalar field in $d=4$ the one-loop vacuum fluctuations diverges
as $\frac{1}{z^{2}}$ if we approach the boundary  at $z=0$ \cite{21}. 
The other term of Eq.(\ref{ult2}) should behave as $\frac{1}{(L-z)^{d-2}}$. To see this 
let us investigate the behavior of the first
integral of $f_2(L,m,d,z)|_{m=0}$ near the boundary at $z=L$.
In order to do this, we make use of
two formulas involving the definition for the Gamma function, 
and also another well known integral representation for the product of
the Gamma function times the Hurwitz zeta function given by 
\begin{equation}
\int_{0}^{\infty} dx \,x^{\mu-1}e^{-\beta x}(\coth
x-1)=2^{1-\mu}\Gamma(\mu)
\zeta(\mu,\frac{\beta}{2}+1)\,\,\,\,Re(\beta)>0,\,\,\,Re(\mu)>1,
\label{I2}
\end{equation}
where $\zeta(z,a)$ is the Hurwitz zeta function
defined by \cite{grads}
\begin{equation}
\zeta(z,a)=\sum_{n=0}^{\infty}\frac{1}{(n+a)^{z}},\,\,\,\,Re(z)>1,
\,\,\,\,\, a \neq 0,-1,-2...
\label{na}
\end{equation}
From the definition of 
the Gamma function and Eq.(\ref{I2}) 
in Eq.(\ref{ult2}) we may write the following closed expression
\begin{eqnarray}
f_{2}(L,m,d,z)|_{m=0}&=&\frac{h(d)}{2L^{d-2}}
\left[2^{2-d}\Gamma(d-2)\left(\zeta(d-2,\frac{z}{L}+1)+
\zeta(d-2,-\frac{z}{L}+1)\right)\right]\nonumber \\
&+& \frac{1}{(2z)^{d-2}}h(d)\Gamma(d-2). 
\label{fim}
\end{eqnarray}
From this last expression and using
the definition of the 
Hurwitz zeta function giving by Eq.(\ref{na}) it is evident
that the regularized $f_{2}(L,m,d,z)|_{m=0}$ has 
two poles of order $(d-2)$, one at $z=0$ and 
another at $z=L$.

To study the massive case,
from the expression given by Eq.(\ref{eq:f2-integrado-en-angulo}) it is possible 
to write $f_2(L,m,d,z)$ in a more convenient way by:
\begin{equation}
f_2(L,m,d,z)=f_{21}(L,m,d,z)+f_{22}(L,m,d,z).
\label{n2}
\end{equation}
where $f_{21}(L,m,d,z)$ and $f_{22}(L,m,d,z)$ are 
\begin{equation}
f_{21}(L,m,d,z)=\frac{1}{2}h(d)
\int_{m}^{\infty}ds(s^2-m^2)^{\frac{d-3}{2}}e^{-2zs},\qquad
\label{n3}
\end{equation}
and
\begin{equation}
f_{22}(L,m,d,z)=\frac{1}{2}h(d)
\int_{m}^{\infty}ds(s^2-m^2)^{\frac{d-3}{2}}(\coth Ls-1)\cosh 2zs.
\label{n4}
\end{equation}
Using an integral representation of the Bessel function of third 
kind or Macdonald's functions it is possible to 
find the following closed expression of $f_{12}(L,m,d,z)$ given by 
\begin{equation}
f_{21}(L,m,d,z)=\frac{1}{2}\frac{1}{(2\sqrt{\pi})^
{d-1}}(\frac{m}{z})^{\frac{d-2}{2}} K_{\frac{d-2}{2}}(2mz).
\label{n5}
\end{equation}
For small $z$ and finite $m$ we have the asymptotic formula 
$K_{\nu}(z)\approx 2^{\nu-1}\Gamma(\nu)z^{-\nu}$, thus for $z\rightarrow 0^{+}$, the 
function $f_{21}(L,m,d,z)$ diverges as $\frac{1}{z^{d-2}}$.
To calculate $f_{22}(L,m,d,z)$ we will use the same method that we used in 
section II. Again note that it 
contains a  power of a binomial. Making use of the generalized 
binomial formula gives
\begin{equation}
(1-\frac{m^{2}}{s^{2}})^{\frac{d-3}{2}}=
\sum_{k=0}^{\infty}(-1)^{k}C^{k}_{\frac{d-3}{2}}(\frac{m}{s})^{2k},
\label{n6}
\end{equation}
and introducing a new variable $u=Ls$ we obtain
\begin{equation}
f_{22}(L,m,d,z)
=\frac{h(d)}{2L^{d-2}}\sum_{k=0}^{\infty}(-1)^{k}C^{k}_{\frac{d-3}{2}}(Lm)^{2k}
\int_{Lm}^{\infty}du\,{}u^{d-3-2k}
(\coth u-1)\cosh(\frac{2zu}{L})
\label{n7}
\end{equation}
Our next step is to show that this result can be expressed in terms
of the Hurwitz zeta function. A natural way to achieve the
proof is to split $f_{22}(L,m,z,d)$ as a sum of two terms
\begin{equation}
f_{22}(L,m,d,z)=f_{22}^{<}(L,m,z,d)+f_{22}^{>}(L,m,z,d),
\label{aa18}
\end{equation}
where
\begin{equation}
f_{22}^{<}(L,m,z,d)=-\frac{1}{4L^{d-2}}
\sum_{k=0}^{k<\frac{d-3}{2}}C^{(1)}(d,k)(Lm)^{2k}
\int_{Lm}^{\infty} du\, u^{d-3-2k} (\coth u
-1)\cosh(\frac{2uz}{L}),
\label{aa19}
\end{equation}
and
\begin{equation}
f_{22}^{>}(L,m,z, d)=-\frac{1}{4L^{d-2}}
\sum_{k\geq \frac{d-3}{2}}^{\infty}C^{(1)}(d,k)(Lm)^{2k}
\int_{Lm}^{\infty} du\, u^{d-3-2k}
(\coth u -1)\cosh(\frac{2uz}{L}).
\label{aa20}
\end{equation}
Here we have introduced
$C^{(1)}(d,k)=(-1)^{k}C_{\frac{d-3}{2}}^{k}h(d)$ and also
$C^{(2)}(d,k)\equiv\frac{\Gamma(d-2-2k)}{2^{d-3-2k}}C^{(1)}(d,k)$
it is possible
to write Eq.(\ref{aa19}) in the following way:
\begin{eqnarray}
f_{22}^{<}(L,m,z,d)&=&-\frac{1}{4L^{d-2}}
\sum_{k=0}^{k<\frac{d-3}{2}}C^{(2)}(d,k)(Lm)^{2k}
\left(\zeta(d-2-2k,-\frac{z}{L}+1)+
\zeta(d-2-2k,\frac{z}{L}+1)\right)\nonumber\\
&&
+\frac{1}{4L^{d-2}}
\sum_{k=0}^{k<\frac{d-3}{2}}
C^{(1)}(d,k)(Lm)^{2k}
\int_{0}^{Lm} du\, u^{d-3-2k}
(\coth u -1)\cosh(\frac{2uz}{L}),
\label{aa21}
\end{eqnarray}
where the singularities of $f_{22}^{<}(L,m,z,d)$ appear at
$z\rightarrow L$. Turning our attention to $f_{22}^{>}(L,m,z,d)$, it is clear that
in the expression above  we see that the surface
divergences are the same as we studied before in the massless case.

We now turn our attention back to the four-point Schwinger function in the 
one-loop approximation.
Introducing new variables as
$u_\pm{}\equiv{}z\pm{}z'$,
the two-point Schwinger function in the tree-level can be split into
\begin{equation}
G_{0}^{(2)}(\vec{\rho},z,z')=G_{+}^{(2)}(\vec{\rho},u_+)+
G_{-}^{(2)}(\vec{\rho},u_-),
\label{split-of-G}
\end{equation}
where making use of the definition of 
$I_{n}(L,m,d,\vec{\rho})$ given by Eq.(\ref{v}) we have 
\begin{equation}
G_{\pm}^{(2)}(\vec{\rho},u_\pm)=
\mp\frac{1}{L}\sum_{n=1}^{\infty}
cos(\frac{n\pi u_\pm}{L})I_{n}(L,m,d,\vec{\rho}).
\label{D-s}
\end{equation}
Before continue, let us 
present a explicit formula of the  
free two-point Schwinger function in terms of Bessel functions. 
Defining an analytic function $g(d)$ by 
\begin{equation}
g(d)=\frac{1}{\sqrt{\pi}(2\pi)^{\frac{d-1}{2}}}
\frac{\Gamma(\frac{d-2}{2})}{\Gamma(\frac{d-3}{2})}, 
\end{equation}
it is possible to show that we can write 
$G_{\pm}^{(2)}(\rho,u_\pm)$ as 
\begin{equation}
G_{\pm}^{(2)}(\rho ,u_\pm)=\mp\frac{g(d)}{\rho^{\frac{d-3}{2}}L}
\sum_{n=1}^\infty{}cos(\frac{n\pi{}u_\pm}{L})
((\frac{n\pi}{L})^2+m^2)^\frac{d-3}{4}
K_{\frac{d-3}{2}}(\rho(m^2+(\frac{n\pi}{L})^2)^{\frac{1}{2}})
\end{equation}
Using Eq.(\ref{split-of-G}) and the  
above formula gives us the explicit expression for the two-point Schwinger 
function in a generic $d$ dimensional Euclidean space confined between two flat paralel 
hyperplanes where we assume Dirichlet-Dirichlet boundary conditions. 
It is hard to use the above expressions for $G_{\pm}^{(2)}(\rho ,u_\pm)$ to 
investigate the analytic structure of the four point function 
$G_{2}^{(4)}(\lambda,\vec{r}_{1},z_{1},\vec{r}_{2},z_{2},\vec{r}_{3},z_{3},\vec{r}_{4},z_{4})$, 
given by Eq.(\ref{nova}), for both the bulk and near the boundaries, Nevertheless
from Eqs.(\ref{g1}) and Eq.(\ref{v}) it is clear that the divergences of 
the four-point 
function in the one-loop approximation appear at coincident points and therefore 
the singular behavior is encoded in the polar part of $M(\lambda,L,m,d)$ given by
\begin{equation}
M(\lambda,L,m,d)=
\lambda^{2}\int d^{d-1}r \int d^{d-1}r^{\prime}
\int_{0}^{L}dz \int_{0}^{L}dz^{\prime} F(\vec{r},\vec{r}',z,z')
(G_{0}^{(2)}(\vec{r}-\vec{{r}^{\prime}},z,z^{\prime}))^2,
\label{amp}
\end{equation}
where $F(\vec{r},\vec{r}',z,z')$ is a regular function. As with the 
one-loop two point function, it is 
not difficult to realize that the above equation has two kinds of singularities, 
those comming from the bulk and those arising from the behavior near the 
surface. As before the behavior in the bulk is as that found in thermal field 
theory and consequently we will only discuss the singularities that arise from 
the boundaries. This 
can be done by studying the polar part of ${\tilde{M}}(\lambda,L,m,d)$ given by  
\begin{equation}
{\tilde{M}}(\lambda,L,m,d)=
\frac{\lambda^{2}}{2}
\int_{0}^{L}dz \int_{0}^{L}dz^{\prime}{\cal{F}}(z,z')
(G_{0}^{(2)}(\vec{0},z,z^{\prime}))^2,
\label{amp2}
\end{equation}
where ${\cal{F}}(z,z')$ is a regular function. Now,  we recall 
that the form of $G_{\pm}^{(2)}(\rho,u_\pm)|_{\rho=0}$ is given by,
\begin{equation}
G_\pm^{(2)}(\rho,u_\pm)|_{\rho=0}=
\mp\frac{1}{(2\pi)^{d-1}L}\sum_{n=1}^{\infty}
cos({\frac{n\pi{u_\pm}}{L}})
\int {d}^{d-1}p
\frac{1}{(\vec{p}^2+m^2+(\frac{n\pi}{L})^2)},
\end{equation}
from where it is not difficult to show that the free correlation function is given by  
\begin{equation}
G_{0}^{(2)}(\rho,z,z')|_{\rho=0}=f_{2}(L,m,d,\frac{u_{-}}{2})-
f_{2}(L,m,d,\frac{u_{+}}{2}).
\end{equation}
For the sake of simplicity we will discuss only the massless case since 
the singularities of the massive case have the same structure as the massless one.
The function 
$f_{2}(L,m,d,\frac{u_{+}}{2})$ is non singular in the bulk, i.e. in the 
interior of the interval $[0,L]$, while $f_{2}(L,m,d,\frac{u_{-}}{2})$ has 
a singularity along the line $z=z^{\prime}$. Indeed, closer inspection shows that 
for $0\leq{}z,z^{\prime}\leq{}L$ the only singularities are those at $u_{+}=0$, $u_{+}=2L$ 
and also $u_{-}=0$. The former two are genuinely boundary singularities (the 
two conditions imply $z,z^{\prime}\rightarrow{}0$ or $z,z^{\prime}\rightarrow{}L$)
while the other comming from $z=z'$ in the whole domain is just the standard 
bulk singularity. In fact, using the structure of the two point 
function and showing just those terms from which singularities might arise, 
one finds that the counterterms for ${\tilde{M}}$ are given by
\begin{equation}
-\mbox{pole}\int_0^Ldz\,\int_0^Ldz^{\prime}[\frac{C_1}{(z+z^\prime)^{d-2}}+
\frac{C_2}{(2L-z-z^\prime)^{d-2}}
+\frac{C_3}{(z-z^\prime)^{d-2}}+...]^{2}.
\end{equation}
where $C_{i}, i=1,..3$ are regular functions 
that do not depend on $z$ or $z'$. From this discussion it is clear 
that in order to render the field theory finite, 
we must introduce surface terms in the action. This is a general statement. For 
any fields that satisfy boundary condition that breaks the translational invariance,
in addition to the usual bulk counterterms, it is sufficient to introduce
surface counterterms in the action to render the theory finite.

\section{Boundary effects and renormalization}

In the last section we present the one-loop renormalization of the 
$\lambda\varphi^{4}$ model, and we considered 
that the field $\varphi(x)$ depends on
$d-1$ unbounded coordinates that we call 
$\vec{r}$ and one bounded coordinate defined in the interval
$[0,L]$. The boundary
conditions on the hyperplanes $z=0$ and $z=L$ are the usual 
Dirichlet-Dirichlet and also Neumann-Neumann boundary conditions. 

In this section we would like to discuss briefly the global approach, used to 
define the Casimir energy associated with any field in the presence of surfaces 
with well defined geometric shape.
The crucial 
conceptual question is the meaning of the renormalized vacuum energy associated 
with any field in the presence of any macroscopic structure that divides the 
space into the internal and the external region.  It is important to keep 
separate different situations. In the case of the parallel plates, the 
region outside the plates is the union of two simple connected domains and both
have the same geometry of the internal region. In this situation  
the Casimir renormalization procedure 
is well defined and the renormalized vacuum energy is unambiguous defined.
In the case of the spherical 
an the cylindrical shell,
the contribution of the exterior modes are not cancelled out in the 
Casimir renormalization
procedure. It is not difficult to understand the origin of the problem, 
as has been extensively discussed in the literature. 
If we are assuming perfectly reflecting boundaries, by the Weyl theorem 
we know that the assymptoptic distribution of eigenvalues of some elliptic differential 
operator is related with the geometric invariants associated with the surface 
where the field satisfy some boundary condition \cite{cou} \cite{babo}. Consequently
in the regularized energy we have divergent 
terms proportional to the volume, area, etc. It the Casimir definition of 
the renormalized vacuum energy it is not possible to cancel the area 
contribution for a generic surface. 
The generalization of the Weyl's expansion can be done investigating the trace
of the heat kernel on a specified manifold with boundary. 
We conclude that the assumption of 
perfect conducting static boundaries with a generic shape introduces new 
problems in order to define the renormalized vacuum energy of a quantum 
system in the presence of these macroscopic objects~\cite{deutsch} \cite{nb} 
\cite{fulling} \cite{milton}. If someone insists in the assumption of perfect 
conducting boundaries there are different ways to solve the problem of infinite 
energy associated with the configuration. 
One is to introduce counterterms concentrated on the 
boundaries, as has been discussed by the authors that use the 
generalized zeta function method \cite{kennedy}. 
A different approach is to smooth out the plate surface 
by a classical potential~\cite{aaa} \cite{cpn}. It is clear that 
the introduction of a classical potential $V(x)$ does not solve the problem of 
surface counterterms since in this situation 
we have to renormalize the potential. A very simple situation is the case of a 
background field where to compute the effective action we have to evaluate the 
the following Fredholm determinant where we are assuming that 
the positive potential is a large quantity.
\begin{equation}  
D(V)=det\,(-\Delta+m^{2}+V(x))(-\Delta+m^{2})^{-1}.
\label{n1}
\end{equation}
For sufficient regular but large $V(x)$ it is possible to show that for $d=4$ a 
counterterm quadratic in $V$ is required in order to eliminate the divergences of the
model \cite{zzz}. Thus the introduction of a 
classical potential $V(x)$ trying to improve the 
unphysical boundary condition does not solve the problem of 
surface counterterms since in this situation 
we have to renormalize the potential.
Instead of smoothing the plates surfaces, a more fruitful approach to avoid 
surface divergences, discussed by Kennedy et al \cite{kennedy} is to treat 
the boundary as a quantum mechanical object. This approach was developed 
recently by Ford and Svaiter \cite{pra} to produce finite values for 
the renormalized $<\varphi^{2}(x)>$ and other quantities that diverge as 
one approaches the classical boundary. 
We would like to stress 
that there will not be any 
surface divergences in a more exact treatment, however 
one can still make the case that surface counterterms are a 
useful phenomenological approach for dealing with the 
apparent surface divergences withouth going 
into the complexity of the more exact approach.

\section{Conclusions}

In this paper we discussed the approach of effective theory 
to perform calculations in field theory in the presence of 
macroscopic structures.
We first assumed the theory of two interacting massive
scalar fields $\varphi_{1}(x)$ and $\varphi_{2}(x)$
with masses
$m_{1}$ and $m_{2}$ satisfying the condition $m_{2}>>m_{1}$.
Integrating out the modes of the field $\varphi_2(x)$
we obtained an effective Lagrangian density 
for $\varphi_1(x)$. In the limit $m_2\rightarrow\infty$ 
the field $\varphi_2(x)$ decouples from $\varphi_1(x)$, 
the only effect of $\varphi_2(x)$ being modifying both 
the value of the renormalized mass 
$m_1$ and the coupling constant of the light field $\varphi_1$.
Thus we considered the $\frac{\lambda}{4!}\varphi^{4}_{1}$
model on a $d$-dimensional Euclidean space, where all
but one of the coordinates are unbounded. Translation
invariance along the bounded
coordinate, $z$, which lies in the interval $[0,L]$,
is broken because
of the boundary conditions (BC's) chosen for the
hyperplanes $z=0$ and
$z=L$. Two different possibilities for these BC's
boundary conditions
are considered: $DD$ and $NN$, where D denotes
Dirichlet and N Neumann, respectively. The renormalization procedure  up to
one-loop order was implemented. The main result of our investigations is 
that in the presence of boundaries where the field satisfies some 
boundary condition, the augmented 
action with surface counterterms can deal with the surface divergences that 
appear in the one-loop Feynman diagrams.

There are several directions for future research 
in field theory in the presence of surfaces, of which we would like to 
emphasize two. The first one is to 
implement the renormalization program 
beyond the one-loop approximation, 
where overlapping divergences 
emerge. The second one is related to the infrared divergences. As we discussed,   
one way to deal with the infrared divergences in the 
case of Neumann-Neumann boundary condition in to perform a resummation of the 
daisy diagrams, although this procedure 
is standard in the study of scalar models at finite temperature,
for systems where the translational invariance is broken, 
it is a open problem how to perform the resummation program. 
Both subjects are under investigation by the authors.

\section{Acknowlegements}

We would like to thank L. H. Ford and L. Bettencourt for useful discussions.
We would also like to acknowledge the hospitality of the Center of
Theoretical Physiscs, Laboratory for Nuclear Science and
Department of Physics of the Massachusetts Institute of
Technology. N.F.S. was partly supported by Conselho Nacional de
Desenvolvimento Cientifico e Tecnol\'ogico do Brazil (CNPq).
M.I.C. was supported by a sabbatical grant from Universidad
Sim\'on Bol\'{\i}var, our visit to C.T.P. was also partially
supported by funds provided by the U.S. Department of Energy
(D.O.E.) under cooperative research agreement DF-FCO2-94ER40810.

\end{document}